\documentclass[sigconf]{acmart}

\PassOptionsToPackage{bookmarks=false}{hyperref}
\hypersetup{bookmarksdepth=-2}

\usepackage[utf8]{inputenc}

\setcopyright{acmcopyright}

\acmDOI{10.475/123_4}

\acmISBN{123-4567-24-567/08/06}

\acmConference[ICSE 2018]{International Conference on Software Engineering}{June 2018}{Gothenburg, Sweden} 
\acmYear{2018}
\copyrightyear{2018}

\acmArticle{4}
\acmPrice{0.00}

\begin{document}
\title{Agile Development for Vulnerable Populations}
\subtitle{Lessons learned and Recommendations}

\author{Marcos Baez}
\affiliation{%
  \institution{University of Trento, Italy}
}
\affiliation{%
\institution{Tomsk Polytechnic University}
}
\email{baez@disi.unitn.it}

\author{Fabio Casati}
\affiliation{%
  \institution{University of Trento, Italy} 
}
\affiliation{%
\institution{Tomsk Polytechnic University}
}
\email{fabio.casati@unitn.it}

\renewcommand{\shortauthors}{Marcos Baez and Fabio Casati.}

\begin{abstract}
In this paper we draw attention to the challenges of managing software projects for vulnerable populations, i.e., people potentially exposed to harm or not capable of protecting their own interests. The focus on human aspects, and particularly, the inclusion of human-centered approaches, has been a popular topic in the software engineering community. We argue, however, that current literature provides little understanding and guidance on how to approach these type of scenarios. Here, we shed some light on the topic by reporting on our experiences in developing innovative solutions for the residential care scenario, outlining potential issues and recommendations.

\end{abstract}


\copyrightyear{2018}
\acmYear{2018}
\setcopyright{acmcopyright}
\acmConference[ICSE-SEIS'18]{40th International Conference on Software
Engineering: Software Track}{May 27-June 3, 2018}{Gothenburg, Sweden}
\acmBooktitle{ICSE-SEIS'18: 40th International Conference on Software
Engineering: Software Track, May 27-June 3, 2018, Gothenburg, Sweden}
\acmPrice{15.00}
\acmDOI{10.1145/3183428.3183439}
\acmISBN{978-1-4503-5661-9/18/05}

\maketitle

\section{Introduction}

The inclusion of human-centered approaches in software engineering has received a lot of attention in recent years \cite{da2011user,salah2014systematic,brhel2015exploring,schon2017agile}. 
Human-centered design (HCD) \cite{dis20099241}, design thinking (DT) \cite{brown2009change} and participatory design (PD) \cite{spinuzzi2005methodology} have been shown to be beneficial for the software design and development process \cite{da2011user}, especially if the designer is aware of their challenges and limitations \cite{salah2014systematic,Bordin2016}.







While there is a growing body of research in exploring user research and agile methods, very little has been done in the area of designing software for \textit{vulnerable populations}. 
The exact definition of the term "vulnerable populations" is the focus of many discussions~\cite{ruof2004vulnerability}. 
Sometimes it is defined vaguely, other times by extension (listing the conditions of users).
In this paper we use it to refer to people potentially exposed to harm or not capable of protecting their own interests.

We argue that software engineering methods - and software engineers - are ill-prepared for addressing this type of projects.
An obvious first consideration is that reasoning on ethics and values tends to be more complex and that each user study requires a very careful design - as well as the need to follow specific guidelines and undergo reviews by an Institutional Review Board (IRB) \cite{national1978belmont}. 
AS we will see the issues go much beyond the incorporation of an ethical approval process (which we found beneficial, besides being appropriate and required).


In this short paper we join the thread of work on the interplay between values and software engineering methods \cite{ferrario2014software,ferrario2016values}. 
We report on our experiences in developing applications for institutionalised older adults over the past years and summarise the lessons we have learned, especially in terms of how we adapted the agile processes we used to follow to cope with the scenarios at hand, and translate the lessons into a corresponding set of recommendations. We hope and trust that this will help teams be more effective when dealing with this scenario and avoid mistakes that can be very costly and hard to recover from.  









\section{Background}

\subsection{Human-centered agile development}

A vast literature has investigated how to combine human-centered approaches with agile methodologies. Systematic literature reviews have focused on the principles user-centered agile development~\cite{brhel2015exploring}, recurring patterns in the integration~\cite{da2011user} and characteristics of stakeholder involvement \cite{schon2017agile}. The most relevant to our discussion is that of Salah et. al~\cite{salah2014systematic}, which summarises the challenges of integrating agile methodologies with HCD.
This review analysed 71 articles on the topic and derived the following challenges: 
i) lack of time for upfront activities, due to the nature of agile development to encourage responsiveness to changes instead of upfront planning, 
ii) conflicts in prioritising UCD and development activities, given the different views on what constitutes progress,   
iii) negative work dynamics arising from potentially competing goals and different communication practices, 
iv) difficulty in organising usability testing and incorporating feedback, due to the time restrictions in agile,  
v) lack of documentation, which creates confusion to UCD practitioners, who are used to record the trace of the design and rationale. 


These reviews, and the works they are based on, provide an insightful perspective on the practices and challenges surrounding human-centered approaches in agile development. However, when vulnerable settings are involved, the challenges become amplified because, as we will see, the timing of access to users, the kind of users we can involve, and the nature and number of iterations is subject to constraints, both self-imposed and imposed by the environment. 

\subsection{Engineering for vulnerable populations}

Efforts have been made in incorporating human-centered approaches in sensitive contexts, and especially in the development of healthcare systems. 
Carroll and Richardson \cite{carroll2016aligning} make a case for the lack of a established framework to guide software developers in identifying requirements in healthcare, and propose integrating design thinking as an entire pre-requirements phase. Another interesting take by Texeira et. al \cite{teixeira2011using} combines more traditional system analysis techniques 
with UCD and PD, which required a facilitator to translate requirements back and forth between stakeholders. The scenario addressed by both works is certainly sensitive, though no actual emphasis is given in dealing with vulnerable populations.

In a similar setting, Kieffer et al. \cite{kieffer2017agile} 
applied agile methods in combination to formative usability 
to the development of an application for patients with diabetes. In reporting the challenges, the authors mention i) the access to users in the medical context, ii) the recruitment process that took about six months in total, iii) the time to get the study protocol validated by an ethical committee, which was four months. On these challenges, the authors reflect that the medical expert should have been involved much earlier in the process. These insights give us a dimension of the practical difficulties in involving vulnerable populations.


Knowledge transfer and communication among multidisciplinary teams is another topic investigated. Weber and Price \cite{weber2016closing} propose a knowledge transfer model between clinicians and software engineers to facilitate the development of healthcare systems. The model is comprised of a knowledge tailoring loop with three main phases: i) monitoring and evaluation of software, in order to collect observation in "real settings" right from the start, ii) identification of problems, involving a qualitative understanding of previous results, and iii) adaptation and tailoring of software, which involves design sessions with lead users and synthesis of results in multidisciplinary teams. While valid in the setting described, using software upfront to gain insights might produce undesired effects, such as the loss of interest by users and stakeholders to continue the collaboration, which is why we believe its applicability with vulnerable populations can be considered limited. 

A more extreme case of team communication was studied by Leonardi et al.
\cite{leonardi2011design}, reporting on the experience of team members with background in HCD and semi-formal requirement engineering. 
The challenge was framed as an inter-cultural dialogue between professionals from different disciplines. The authors stress the importance of mutual learning, especially via the definition of a shared dictionary to bridge the gap between the disciplines.

Speedplay \cite{ferrario2014software} is a software project management framework that integrates action research, participatory design and agile development, to approach relatively small projects targeting specific community needs. It is particularly targeted at multi-disciplinary projects seeking social innovation, where the community, researchers and engineers work actively together. The process model is comprised of four main steps: prepare, design, build and sustain, and it is characterised by slower cycles at the beginning followed by faster paced cycles by the end of the project. The model also promotes mutual learning while assigning responsibilities based on skills.
An application of Speedplay is presented by Simm et al. \cite{simm2014prototyping} in the context of a tool for anxiety management for adults with high functioning autism. In working with this vulnerable population, the authors mention participants reacting poorly to changes and fluctuating participation as some of the challenges to an agile research and development.





While we found many of the guidelines from the literature quite useful and we recognised the same challenges, in our experience we have stumbled upon difficulties and derived insights that we have not seen discussed deeply and uncovered several aspects that have not emerged yet. 
We discuss them in the following.


\section{The case of residential care}
We describe our findings based on a joint university-industry project aiming at designing a set of innovative solutions for the residential care scenario focused at increasing the emotional well-being of residents, staff, and family members and at facilitating interactions. 
We also build on previous experience \cite{fiore2017understanding} studying analogous issues in the pediatric palliative care case.  
In this section we describe the context and give an overview of the project setup.

Transitioning to long-term residential care is one of the more difficult moments in the life of an older adult and their family \cite{lee2002review}. 
This is complicated by the perceived negative social view on residential care that sees institutionalisation as a failure by family members \cite{ryan2000nursing}, 
due to cultural stereotypes about care systems, 
resulting in a sense of guilt, loss and abandonment in the family, as well as a challenging work environment for care professionals. 


The relationship between family members and professional caregivers is not without tensions.
From the family members side, communication is challenged by a perceived lack of meaningful, timely and understandable information \cite{fiore2017understanding}. 
Professional caregivers instead consider "dealing" with family members as part of the job, but can see communications as potentially problematic. This can lead to professionals avoiding  family members and vague communications \cite{hertzberg2003relatives}.
Another critical point is the collaboration between these actors. Though collaboration from family members is in principle welcomed by professionals, this is not always translated into practice as traditional care models are not designed for full collaboration \cite{haesler2004constructive}. 
Family members can also be overly demanding, having unrealistic expectations about what the professionals have to do, and even taking issue in care practices \cite{vinton1998intervening}.
Finally, as we experienced, NH staff members work at full capacity and in rather stressful conditions, both in terms of helping residents and in terms of managing family members demands and expectations.
This is the scenario we were set to work with: very frail participants, emotionally demanding, and possibly conflictive.



We followed a mix of agile and human-centered approaches, iterating on the following three main phases: product discovery, development and validation. Agile methodologies do not address the product discovery phase \cite{brhel2015exploring} but the need for a dedicated phase is evident in human-centered approaches and the software processes that incorporate them (e.g. \cite{ferrario2014software}). 

As a result of the process we identified and developed three IT-based solutions: i) a reminiscence-based tool to stimulate social interactions between family members and residents, ii) a personalised magazine to build a sense of community and stimulate conversations, and iii) a communication and collaboration tool to facilitate information sharing and family involvement. Describing the tools its outside the scope of this paper, and the interested reader is referred to \cite{fiore2017understanding,ibarra2017stimulating,caforio2017viability}. 
\section{Issues and Recommendations}
We now list challenges we had to face due to working with vulnerable subjects and our recommendations. 

\textbf{Iterations are limited, errors are costly.}
Agile processes allow us to iterate often with users and to correct course of actions as needed. This is often done by pushing "software probes" at early phases (e.g., \cite{weber2016closing}).
In our scenario, we found that the number of possible design iterations are limited: There is a relatively small number of institutions willing at the start to go through an adventure with you, the personnel is often under stress, residents and family members face loads of challenges. 
Even those who are enthusiastic at the beginning can become less cooperative (or simply less available) as time goes by. Furthermore, while learning from errors is a positive aspect in agile methods, continuous changes can be disruptive to a population not used to change \cite{simm2014prototyping} and can harm their interest to be involved and participate. 
More importantly, errors in dealing with vulnerable populations can damage the trust, and that is something very difficult to recover from \cite{mara2013ethics}. Here an "error" can be simply giving a hint of suspicion that the system you are building goes in a direction that does not fit the needs of the individual you are speaking with, even if that is not the case. Once this happens, even reassurances that their feedback will be taken into account might not have the hoped effect.

Something we found useful and highly recommend here is to observe users and analyse the relation between stakeholders and technology through the lenses of \emph{appropriation} \cite{dourish2003appropriation}, which is often revealing desired technology features that are satisfied via other means today.
Furthermore, we recommend spending more time at the start assessing assess to participants, and specifically identifying participants for which we can have an "agile" style of access and interactions, and participants for which  i) we have reduced access and ii) errors are sensitive.  For the latter categories, we recommend deeper studies before introducing software probes or mockups.


The need for going through \textbf{ethical approval processes} impacts the process in many ways (including positive ones).
Obtaining an approval requires time, both to write the necessary documentation and to go through the approval process, which may involve one or more entities. In our case we went through both University and NH committees. Timescales here are typically of 1-3 months, depending on frequency of committee meetings and on whether clarifications are requested. 
This timescale is already beyond any modern agile standard.
However the process is also an opportunity to carefully think and design the study protocol and receive suggestions. Since we have to be very conservative with iterations and user access, this step is actually helpful especially for a team with an agile mentality which might be tempted to reduce planning and have a bias towards action.
A related aspect is the need to re-assess ethical considerations in an "agile" settings. This not only requires continuous adjustments in the research practice \cite{rashid2015managing}, but can also result in further changes to initial study protocols.

Therefore, our recommendation here is to include in the approval process a structured plan of actions, carefully considering and anticipating possible outcomes and designing subsequent study steps accordingly, as opposed to iterating in an agile way. 
This is both to solicit more informed feedback, but also to limit further requests to the ethical committee to hopefully minor modifications to a plan (if at all), which typically result in faster feedback.

Related to appropriation and ethics is the issue of 
\textbf{workarounds} that people (and specifically staff members) take, very often with the intent to help people in need.
We found this to happen in all scenarios involving vulnerable subjects, beyond the case of institutionalised older adults.
Observing workarounds is  very useful in design, but in sensitive settings such as the one investigated, the "creativity", ingenuity, or simply the commitment and dedication of participants might be perceived as deviations from procedures and sometimes even from the law. 
In this sense, ethical considerations of reporting or incorporating such learnings in the design arise. 
What we recommend here is to ensure that the team is aware and follows recommended ethical guidelines and practices for these cases \cite{mara2013ethics}. This is something that even people trained in user studies may not be familiar with, and an error here may cost people their job.


\textbf{Participant involvement.} From the perspective of participant involvement \cite{radermacher2006participatory}, our approach qualifies as research-initiated, with shared decisions with the industry partner, and nursing homes actors being consulted and informed. This puts us halfway between traditional and more participatory approaches \cite{ferrario2014software}. 
Incorporating a human-centered approach proved to be successful in identifying needs and materialising technology concepts, but   limited in testing more forward thinking features. In a scenario where technology should also follow regulations and (sometimes anticipate) changes in policies, following a full community-driven or human-centered approach is not always feasible \cite{norman2005human}. 
%
Thus, rather than taking a decision a priori, we recommend to base the decision on the level of involvement according to the design goals, potential bias, conflictive views, and ethical considerations.

\textbf{Personas} are designed to evoke emotional responses, creating empathy and keeping the team focused on the target users. 
In our work we had two challenges related to personas: one is access. Even if it is rather easy to identify personas and in general to cluster users into characteristics and needs, 
access to personas in different groups is hard to organise for many reasons.
The other is that sometimes people had strong feelings about the party they interact, which can result in colorful or emotionally provoking personas.
We therefore recommend i) to tone down description of personas when discussing with end users and ii) carefully assess at the start of the project 
(or after the personas have been identified) the access to different personas and prioritise requests: do not expect to manage to have access to all personas.

\textbf{Varying feedback.} Response bias is a widely studied behavior. 
With vulnerable subjects, we found the problem to be exacerbated. 
It manifested itself particularly when staff members, in high level discussions, reported an open attitude towards technology supporting staff-FM interactions (in accordance with the management and the political decision makers). 
However, when we drilled down into understanding how the communication should take place in very specific scenarios, we observed a certain resistance in by staff members related to some potential features. 
This occurred despite the research team being competent and trained in how to run studies with vulnerable subjects. 
In retrospect, drilling down to details earlier would have avoided us to work on features that are unlikely to make it in the system. 
We recommend therefore to identify features that might be contested and to drill down on them early, by providing concrete and realistic examples to validate acceptance.

\textbf{Multidisciplinary teams} are essential but difficult to manage. Our core team was comprised of a multidisciplinary group of:
i) sociologists with background in participatory action research and qualitative research methods, 
ii) researchers with background in software engineering and human-computer interaction, 
iii) product managers with experience in the healthcare sector, and iv) cognitive scientists and psychologists with competence on interactions and stress. 
Interactions with vulnerable populations require empathy,  soft skills, experiences in designing studies in a way that is mindful of biases, and the ability to avoid putting the participants in an uncomfortable situation. This is hardly something that can be learned with a crash course.
Nonetheless, involving software engineers in informal visits -- and user studies when possible -- has proven useful in our experience in creating empathy and having a more realistic view of the context, which was later useful when incorporating and discussing the lessons learned.

The communication and collaboration in the team was facilitated by the cross-functional team members with experience in software engineering and human factors. This setup is known to facilitate the integration of design and development \cite{brhel2015exploring}, minimising the need for resorting to, for example, translations between team members \cite{leonardi2011design}. This was further facilitated by the use of scenarios, personas and mockups that were concrete materialisations of lessons learned. 
However, coordinating efforts towards activities that would maximise requirement elicitation has been more challenging. 
We see this as a consequence of competing views on what qualifies as useful insights among sociologists and engineers. Cross-functional members were fundamental in mediating these differences. Their importance cannot be overestimated and we feel that the lack of such competences can jeopardise the design effort.




In summary, what we take home is the need to mix agile approaches with waterfall concepts. 
With some participants we can follow and iterate with short-lived design or development sprints, while with vulnerable populations we execute much longer sprints, characterised by a thorough design process that anticipates possible alternatives as opposed to designing them iteratively. 
Participants involvement needs the same flexibility: with some users observation only is appropriate, with others we can leverage PD, with others again we can follow traditional user research.
We find that the challenge of design lies therefore not in the choice of a specific process and model for the project, but in identifying which participants and which tasks are suited for a given process and design approach.



\begin{acks}
This project has received funding from the EU Horizon 2020 research and innovation programme under the Marie Sk\l{}odowska-Curie grant agreement No 690962, for the studies run outside the EU. This work was also supported by the ``Collegamenti'' project funded by the Province of Trento (l.p. n.6-December 13rd 1999), for the studies run in Italy.
\end{acks}

\bibliographystyle{ACM-Reference-Format}
\bibliography{references,domain} 

\end{document}